\documentclass[conference]{IEEEtran}
\IEEEoverridecommandlockouts
\usepackage{cite}
\usepackage{amsmath,amssymb,amsfonts}
\usepackage{graphicx}
\usepackage{textcomp}
\usepackage{xcolor}
\usepackage{url}
\usepackage{booktabs}
\usepackage{pifont}
\usepackage{algorithm}
\usepackage{algpseudocode}

\newcommand{\callout}[1]{\ding{\numexpr181+#1\relax}}

\begin{document}

\title{WeaveData: A Multimodal Data Analysis System with Self-Critiquing and Self-Evolving LLM Plans}

\author{
\IEEEauthorblockN{Min Jia$^\dag$$^\P$, Shihao Zhou$^{\S}$, Jun-Peng Zhu$^\dag$$^\P$, Peng Cai$^{\S}$, Kai Xu$^\P$, Chao Zhang$^\ddag$\\
Li Li$^\P$, Aoying Zhou$^\S$, Heng Long$^\P$, Qiu Cui$^\P$, Liu Tang$^\P$, Qi Liu$^\P$ }
\IEEEauthorblockA{\textit{$^\dag$Northwest A\&F University, $^\S$East China Normal University, $^\ddag$Renmin University of China, $^\P$PingCAP}}
\IEEEauthorblockA{\{mjia, zjp.dase\}@nwafu.edu.cn, shihao.zhou@stu.ecnu.edu.cn, \{pcai,ayzhou\}@dase.ecnu.edu.cn \\
cycchao@ruc.edu.cn, \{xukai,lili,lh,cuiqiu,tl,liuqi\}@pingcap.com
}
}

\maketitle

\begin{abstract}
Multimodal data analysis, which answers questions over relational tables, text, and images, has attracted growing attention in the data management community.
Large language models (LLMs) enable such analysis in natural language by generating analysis plans over relational and semantic operators.
However, LLM-generated plans are error-prone: a plan may silently compute something other than what was asked, fail during execution, or return a result that misses the question.
This paper presents WeaveData, a multimodal data analysis system with self-critiquing and self-evolving LLM plans.
First, WeaveData generates a typed logical plan for each question and critiques it step by step before execution, and it checks the executed result against the question afterwards.
Second, WeaveData evolves a plan that fails or misses the question: it diagnoses the failure with the actual data, reuses the results that remain valid, and accumulates planning experience for later questions.
Third, WeaveData grounds planning in a metadata knowledge graph of all modalities, clarifies ambiguous questions with the user, and backs every model judgment with evidence in an interactive notebook.
We demonstrate WeaveData on two public multimodal datasets.
\end{abstract}

\begin{figure*}[t]
\centering
\includegraphics[width=\textwidth]{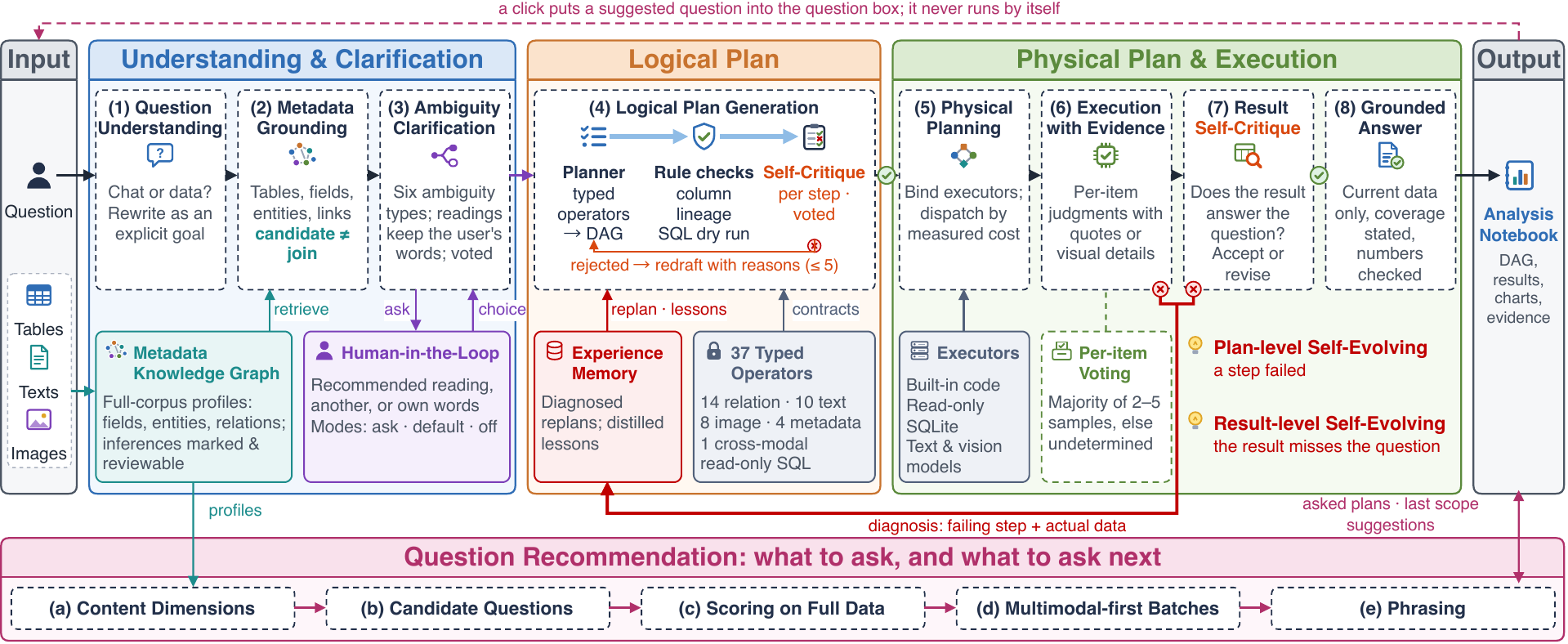}
\caption{Architecture of WeaveData.}
\vspace{-0.3cm}
\label{fig:architecture}
\end{figure*}

\section{Introduction}

Real-world data increasingly combines relational tables with text and images.
A real-estate listing, for example, has a table row with its price and location, a textual description, and several photos, and an analyst may ask ``Which listings near MIT look modern with lots of natural light in their photos?''~\cite{palimpzest}
Answering it requires relational processing and semantic judgments over text and images within one analysis, for which large language models (LLMs) provide new opportunities.
For relational data, LLM-based systems such as Chat2Query~\cite{chat2query} and TiInsight~\cite{tiinsight} already automate exploratory analysis in natural language (NL).
For multimodal data, semantic operator systems such as LOTUS~\cite{lotus}, Palimpzest~\cite{palimpzest}, and ThalamusDB~\cite{thalamusdb} evaluate LLM-based filters, joins, and extractions inside queries, and LLM-based planners such as CAESURA~\cite{caesura}, AOP~\cite{aop}, Unify~\cite{unify}, and Nirvana~\cite{nirvana} translate an NL question into a plan of relational and semantic operators.
In these systems, the plan generated by the LLM determines what is computed, and hence whether the answer is correct.
In our experience with public multimodal benchmarks~\cite{sembench,palimpzest,nirvana}, LLM-based planning still faces three limitations.

\textbf{(1) Plans that silently answer a different question.}
LLM-generated plans often look plausible yet compute something other than what was asked: we observed plans that cut rows tied at the last rank with \texttt{LIMIT}, replaced a semantic judgment over text with a substring match, or counted items that the model could not judge as negative answers.
Such errors raise no exception, and finding them requires inspecting plans with hundreds of model calls.

\textbf{(2) Plans that cannot recover from failures.}
A plan may also fail at run time, e.g., on a join key that is not unique, or return a result that misses part of the question.
A one-shot planner then stops or returns the incomplete result; the analyst must rephrase the question and pay for all model calls again, and the same mistake recurs in later questions.

\textbf{(3) Ambiguous questions over loosely connected sources.}
Multimodal questions are often ambiguous: each listing above has three photos, and ``look modern in their photos'' may require one or all of them to qualify, which returns different listings on the Palimpzest real-estate data~\cite{palimpzest}.
Sources also often share no key, e.g., airline routes and logo images in the MMQA scenario of SemBench~\cite{sembench} can only be linked by what the images show.

To address these limitations, we present WeaveData, a multimodal data analysis system with self-critiquing and self-evolving LLM plans.
WeaveData introduces the following features.
To begin with, WeaveData critiques every plan before it runs, with deterministic checks and an LLM self-critique that verifies the preconditions of every step in execution order, and checks the executed result against the question afterwards.
Second, WeaveData evolves plans that fail or miss the question: it diagnoses the failure with the actual data, reuses valid intermediate results, and distills each successful repair into a lesson for later questions.
Third, WeaveData grounds planning in a metadata knowledge graph of all modalities, asks the user to resolve ambiguities, and links sources without shared keys by their content.
Finally, WeaveData backs every model judgment with a quote or a visual detail and exposes the whole analysis in an interactive notebook.

Through this demo, ICDE attendees can experience WeaveData in action.
After selecting a dataset, attendees can (1) explore its metadata knowledge graph and recommended questions, (2) ask a question in NL and resolve its ambiguity by choosing among ranked interpretations, (3) watch WeaveData draft, critique, execute, and repair the plan, and (4) inspect the evidence behind every intermediate result.
We demonstrate WeaveData on two public multimodal datasets: the MMQA data of SemBench~\cite{sembench} and the real-estate data of Palimpzest~\cite{palimpzest}.
The system is available at \textcolor{blue}{\url{https://github.com/DASE-iDDS/weavedata}}, and the demo video is available at \textcolor{blue}{\url{https://youtu.be/MQXFPBQMFEk}}.

\vspace{-0.3cm}
\section{System Overview}

Fig.~\ref{fig:architecture} illustrates the architecture of WeaveData: a question flows through eight steps in three stages, and a recommendation module suggests what to ask.

\ding{172} \underline{\textbf{Metadata Knowledge Graph.}}
WeaveData profiles every source in advance: column statistics and key references for tables, and, for every text and image, the topics, entities, objects, and subject--predicate--object relations found by the LLM, each backed by a quote or a visible detail.
The profiles form a knowledge graph across modalities, in which model-inferred relations remain candidates until a user confirms them.

\ding{173} \underline{\textbf{Understanding and Clarification (steps 1--3).}}
WeaveData rewrites the question into an explicit goal and retrieves the relevant sources, fields, and relations from the graph.
If the question admits interpretations that lead to different answers, e.g., whether one or all photos of a listing must qualify, WeaveData asks the user to choose one or to state the intent, at most three times.
Interpretations are ranked by how few assumptions they add to the user's words, how well the data supports them, and their cost, never by their results.

\ding{174} \underline{\textbf{Self-Critiquing Logical Planning (step 4).}}
The LLM planner composes a typed logical plan, and every draft is checked by rules and critiqued step by step before it may run (Section~\ref{sec:critique}).

\ding{175} \underline{\textbf{Physical Planning and Execution (steps 5--6).}}
WeaveData binds each logical operator to a trusted executor, i.e., built-in code, read-only SQLite, or a text, vision, or comparison model, and dispatches independent branches by their measured cost.
Except for summarizing a text collection, model-based operators judge one item at a time and back each judgment with a quote or a visible detail; disagreeing samples are settled by majority vote, and an item that cannot be judged stays unknown rather than negative.
Sources without a shared key are linked by matching items with table rows by content.

\ding{176} \underline{\textbf{Result Self-Critique and Self-Evolving (steps 7--8).}}
A critic checks whether the executed result answers the question, and a plan that fails or misses the question evolves (Section~\ref{sec:evolving}).
The final answer uses only executed results, and every number and claim in it is checked against them.

\ding{177} \underline{\textbf{Question Recommendation.}}
WeaveData recommends questions computed from the data and its profiles, prefers cross-modal ones, and labels each with the number of items it needs to read.

\vspace{-0.3cm}
\section{Self-Critiquing Planning}
\label{sec:critique}

LLM planners produce plausible plans that may be wrong in subtle ways.
WeaveData critiques each plan twice: before execution to catch errors in the plan itself, and after execution to check whether the result answers the question.

\underline{\textbf{Typed Logical Plans.}}
A logical plan is a DAG of typed operators.
Each operator declares its input ports, parameters, and output type, which is a table, a set of per-item judgments with their evidence, a text collection, or an image collection.
The 37 operators cover relational processing (e.g., filter, aggregate, and read-only SQL), per-item extraction and classification over text and images, summaries of whole text collections, content-based matching between items and table rows, and metadata exploration.
Because the planner can only compose registered operators, every plan can be checked before execution, and the only model-written code that runs is read-only SQL.

\underline{\textbf{Rule-Based Checks.}}
Each draft first passes deterministic checks that need no LLM call.
Contract checks reject unknown operators, missing parameters, and type mismatches between ports.
Semantic checks reject patterns that silently change the question, such as a \texttt{LIMIT} that cuts ties at the last rank or a substring match in place of a semantic judgment.
Column lineage derives the columns produced by every step, and each SQL query is dry-run on an empty table with its input columns, so a reference to a missing column is caught before any model is called.

\underline{\textbf{Step-by-Step Self-Critique.}}
Rules cannot tell whether a plan means what the user asked.
WeaveData therefore has the LLM critique the plan against a domain definition that specifies the preconditions and effects of every operator.
Walking through the plan in execution order, the critic states each step's preconditions, verifies them against the current state of its inputs (e.g., which columns exist and what one row represents), and derives the resulting state.
Finally, it checks whether the outputs answer every part of the goal and returns one of three verdicts: \emph{the plan is correct}, \emph{the plan is wrong} with its first failing step, or \emph{goal not reached}.
The domain definition also requires faithfulness to the question: a semantic step must state the user's criterion as worded, without widening or narrowing it.
A rejected draft is returned to the planner together with the critique, up to five drafts.
Since a wrong rejection can steer the planner away from a correct plan, a rejection takes effect only if a majority of up to three critiques agree.
When several drafts fail, WeaveData asks whether the operators can carry out the goal at all and, if not, explains why instead of producing a plan that answers something else.

\underline{\textbf{Result Self-Critique.}}
After execution, a second critic examines the actual outputs and the evidence of every step, and decides whether to accept the result, revise the plan, or stop.
It separates deficiencies of the plan from gaps in the data: a missing image or an item that the model cannot judge is reported to the user rather than treated as an error of the plan.

\begin{figure*}[t]
\centering
\includegraphics[width=\textwidth]{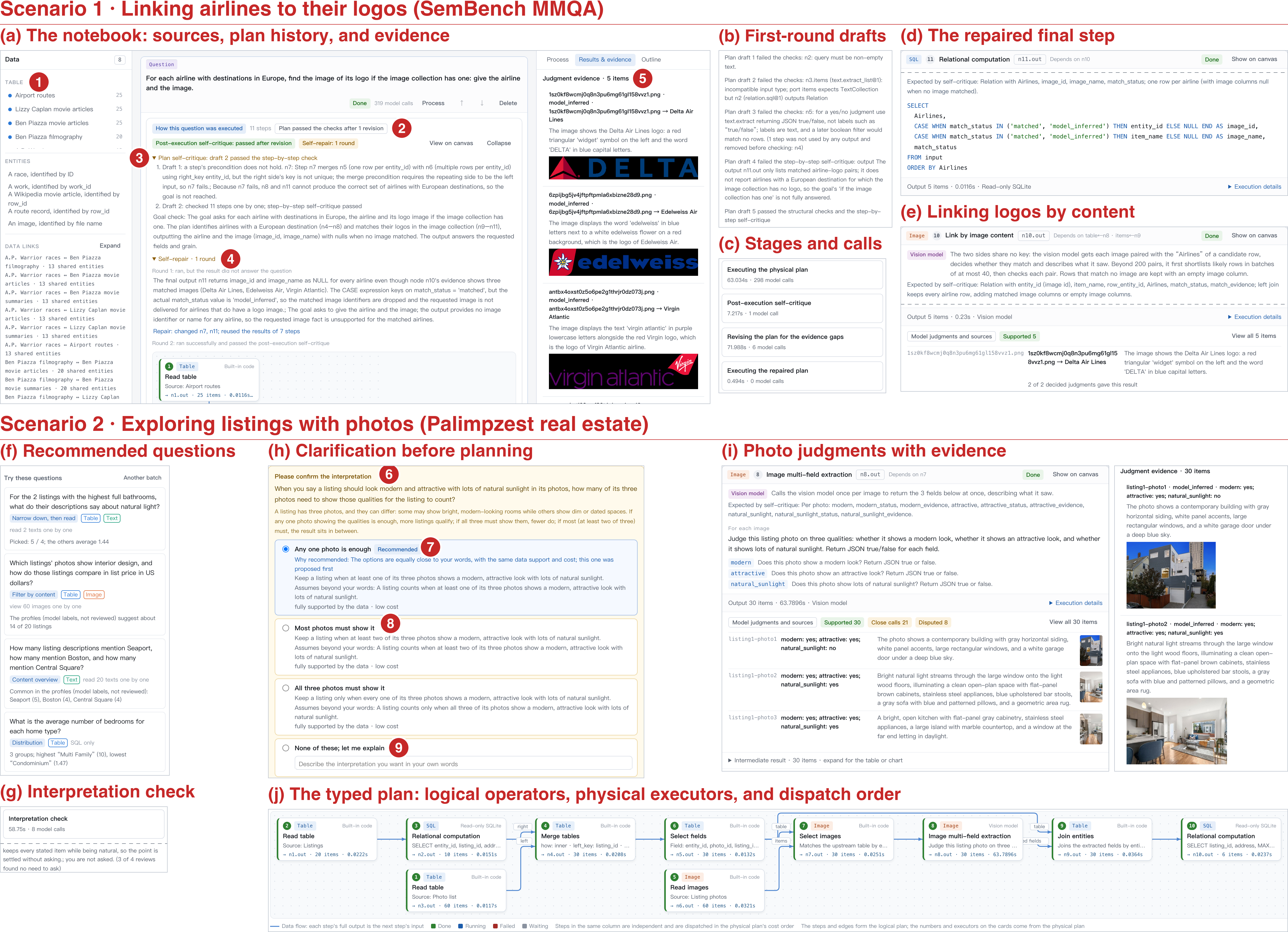}
\caption{WeaveData in the two demonstration scenarios; the annotations are in red. \callout{1}~Sources and the links found between them, \callout{2}~how the question was executed, \callout{3}~self-critique of the plan drafts, \callout{4}~self-repair after the result critique, \callout{5}~judgment evidence, \callout{6}~a clarification question and why it matters, \callout{7}~the recommended interpretation and its reason, \callout{8}~other interpretations, and \callout{9}~an answer in the user's own words.}
\vspace{-0.3cm}
\label{fig:demo}
\end{figure*}

\vspace{-0.3cm}
\section{Self-Evolving Planning}
\label{sec:evolving}

A plan that passes the self-critique can still fall short at run time: a join key that looks unique may repeat in the data, or the result may miss part of the question.
Instead of stopping, WeaveData answers a question in at most four rounds (Algorithm~\ref{alg:evolve}): each round executes the current plan, reusing valid earlier results, and then accepts the result or evolves the plan, which passes all checks of Section~\ref{sec:critique} again.

\begin{algorithm}[t]
\caption{Self-evolving planning of a question $q$}
\label{alg:evolve}
\footnotesize
\begin{algorithmic}[1]
\State $P \gets$ \Call{CritiquedPlan}{$q$, lessons} \Comment{Section~\ref{sec:critique}}
\For{round $r = 1, \dots, 4$}
  \State $R \gets$ \Call{Execute}{$P$}, reusing valid earlier results
  \If{a step of $P$ failed}
    \State $d \gets$ \Call{Diagnose}{first failed step}
    \State \algorithmicif\ no plan can fix $d$ \algorithmicthen\ \Return $d$ with an explanation
    \State $F \gets$ ($d$, repair hint, data facts)
  \Else
    \State $v \gets$ \Call{ResultCritique}{$q$, $P$, $R$}
    \State \algorithmicif\ $v = \textit{accept}$ \algorithmicthen\ distill a lesson if $r > 1$; \Return \Call{Answer}{$R$}
    \State \algorithmicif\ $v = \textit{stop}$ \algorithmicthen\ \Return $R$ with the critic's issues
    \State $F \gets$ (critic's issues, data facts)
  \EndIf
  \State \algorithmicif\ $r < 4$ \algorithmicthen\ $P \gets$ \Call{CritiquedPlan}{$q$, lessons, $P$, $F$}
\EndFor
\State \Return $R$ with $F$
\end{algorithmic}
\end{algorithm}

\underline{\textbf{Plan-Level Evolution.}}
When a step fails, WeaveData uses its error code to decide whether a new plan can fix the failure.
Fixable failures are those a different plan can avoid, e.g., an SQL query that does not run, a missing column, a type mismatch, a non-unique join key, or a result of more than 10,000 rows.
For each of them, WeaveData sends the planner the failed step and a repair hint, e.g., to aggregate the input to one row per key before a join.
Failures caused outside the plan, such as an unavailable model service, an exhausted call budget, or a source that changed during the analysis, cannot be fixed by replanning, so WeaveData stops and explains them.
An error without a type is treated as an internal error rather than a plan error, so that WeaveData does not learn a false lesson from it.
A failing SQL query is first repaired in place: the LLM may only correct its syntax and column references, and a separate critique accepts the correction only if the query keeps its conditions and meaning.

\underline{\textbf{Result-Level Evolution.}}
When the result critique decides to revise, the deficiencies it names, e.g., that requested images are missing from the output, become the feedback for the next plan.

\underline{\textbf{Data-Aware Feedback.}}
A planner that only reads an error message tends to return the same plan.
The feedback therefore includes the prior plan, the steps that succeeded, and data facts that show where the plan went wrong: for a step that returned no rows from a non-empty input, the actual values and types of the columns it compared (e.g., the text \texttt{"true"} where a filter expects the number \texttt{1}); for a failed step, its input columns and first rows.
The planner must change the failing step and keep the steps that need no change; if it returns the prior plan unchanged, WeaveData points out the remaining problem and asks once more before it stops.

\underline{\textbf{Reusing Valid Results.}}
WeaveData identifies each step by a lineage signature, i.e., a hash of its operator, its parameters, and the signatures of its inputs.
A step of the evolved plan reuses the result of an earlier successful step with the same signature if all its inputs are reused as well, so renumbered steps are still recognized.
Moreover, each per-item model judgment is cached under a fingerprint of the operator, the instruction, the item's content, and the model, so a step that runs again pays only for the items it has not judged yet.

\underline{\textbf{Experience Memory.}}
When a question succeeds after at least one repair, the LLM distills the failed rounds and the repaired plan into a lesson: a trigger describing the data situation, what a plan should do, and the operators involved, without names or values from the question; a one-off slip yields no lesson.
For a new question, WeaveData ranks the active lessons of the same data and the general ones by source overlap, similarity to the question, and track record, and passes the top ones to the planner.
The planner must list the lessons it applies, and a lesson is credited with a success when its round passes the result critique and with a failure otherwise.
A lesson learned on some data becomes general after three successes or a success on other data, and it is retired after at least two failures that outnumber its successes.
Because credit follows outcomes rather than causes, lessons remain hints: all checks still apply, users can retire and restore lessons, and evaluations turn the memory off.

\section{Demonstration Scenarios}
\label{sec:demo}

\vspace{-0.2cm}
\subsection{The WeaveData Notebook}

After an attendee opens a dataset, WeaveData shows an analysis notebook (Fig.~\ref{fig:demo}(a)).
The left pane lists the sources with their entities and the links found between them (\callout{1}).
In the center, each question is followed by a summary of how it was executed (\callout{2}): the number of steps, the revisions the plan needed to pass the checks, the verdict of the result self-critique, and the rounds of self-repair.
Attendees can expand the self-critique of the drafts (\callout{3}) and the history of the self-repair (\callout{4}); below them follow the plan DAG and one cell per step with its executor, the state that the self-critique expected, its result, and the evidence of its model judgments.
The right pane shows the planning log (Fig.~\ref{fig:demo}(b)), the stages of the analysis with their time and model calls (Fig.~\ref{fig:demo}(c)), the results and evidence of a selected step (\callout{5}), and recommended questions (Fig.~\ref{fig:demo}(f)); new questions are asked in NL at the bottom.

\subsection{Scenario 1: Linking Airlines to Their Logos}

On the MMQA data of SemBench~\cite{sembench}, we ask ``For each airline with destinations in Europe, find the image of its logo if the image collection has one: give the airline and the image.''

\textbf{Step 1: Self-critiquing planning.}
Grounded in the metadata, the planner first explains which sources it needs: no table identifies airlines as entities, so the images must be linked to the airline names by their content.
Rule checks then reject three drafts without any model call (Fig.~\ref{fig:demo}(b)): one leaves an SQL query empty, one feeds a table into a text operator, and one returns a yes/no judgment as text labels that a later boolean filter would never match.
The step-by-step self-critique rejects the fourth draft because its output lists only the airlines with a matched logo, whereas the question asks about every airline with a destination in Europe.
The fifth draft passes: the text model decides whether each airline's destination list includes a destination in Europe, and the vision model matches the 25 images with the airline names by their content.
Every step cell records the state that the self-critique expected, e.g., that the matching step keeps every airline row and adds empty image columns when no logo matches (Fig.~\ref{fig:demo}(e)).

\textbf{Step 2: Self-evolving planning.}
The first execution takes 298 model calls (Fig.~\ref{fig:demo}(c)).
The result critique then finds that the output contains no image although the matching step found logos, and it names the cause from the actual data: the final SQL step keeps an image only if its match status is \texttt{matched}, whereas the actual status is \texttt{model\_inferred} (\callout{4}).
The evolved plan changes two steps and reuses the results of seven (\callout{4}); its first draft is rejected by the self-critique because its merge step would join on a key that repeats on the right side, and the second draft passes (\callout{3}).
The repaired final step also keeps the images whose status is \texttt{model\_inferred} (Fig.~\ref{fig:demo}(d)).
Revising the plan takes six model calls, and executing the repaired plan takes 0.5\,s and no model call: the logo-matching step runs again but finds all its judgments in the cache (Fig.~\ref{fig:demo}(c)).

\textbf{Step 3: Evidence.}
The answer lists five airlines with their logo images, in agreement with SemBench's ground truth, and states how the result was computed and how many judgments were inferred or unknown.
Each match carries the detail that the vision model saw and its votes, e.g., ``a red triangular `widget' symbol on the left and the word `DELTA' in blue capital letters'', given by 2 of 2 decided judgments (Fig.~\ref{fig:demo}(e)), and the right pane shows the matched images (\callout{5}).

\subsection{Scenario 2: Exploring Listings with Photos}

On the real-estate data of Palimpzest~\cite{palimpzest}, WeaveData recommends questions before the first one is asked; they are computed from the data and its profiles, and each is labeled with the modalities it combines, the number of texts or images it must read, and the data fact that motivates it (Fig.~\ref{fig:demo}(f)).
We ask ``Which listings within two miles of MIT, priced above \$600,000 and up to \$2,000,000, look modern and attractive with lots of natural sunlight in their photos? Give the listing\_id and address.''

\textbf{Step 1: Clarification.}
Since the question does not say how many of a listing's three photos must qualify, WeaveData pauses and asks (Fig.~\ref{fig:demo}(h)): it explains why the choice changes the answer (\callout{6}), recommends that any one photo is enough and gives the reason (\callout{7}), lists the other interpretations with the assumptions they add, their data support, and their cost (\callout{8}), and accepts an answer in the user's own words (\callout{9}).
The interpretation check behind this question takes eight model calls, and it drops another candidate question because three of four reviews find it unnecessary (Fig.~\ref{fig:demo}(g)).
A second round asks whether a photo must show the three qualities together; we accept the recommended interpretations, which stay attached to the question.

\textbf{Step 2: Planning and execution with evidence.}
The typed plan passes the checks at its first draft; its cards show the operator, the executor, and the dispatch order that the physical plan assigns to each step (Fig.~\ref{fig:demo}(j)).
The plan narrows the listings to ten with SQL, has the vision model judge whether each of their 30 photos looks modern, attractive, and sunlit, with the detail it saw (Fig.~\ref{fig:demo}(i)), and keeps the six listings that have a photo showing all three qualities.
The vision step flags 21 of the 30 judgments as close calls, where the photo leaned only weakly or the repeated judgments differed, and 8 as disputed, which record the majority, so attendees know which photos to inspect.

\section{Conclusion}

We presented WeaveData, a multimodal data analysis system whose LLM-written plans critique themselves before execution and evolve when they fail or miss the question, reusing valid results and accumulating experience. The demonstration shows these capabilities end to end on two public multimodal datasets.

\bibliographystyle{IEEEtran}
\bibliography{reference}

@article{lotus,
  author  = {Liana Patel and Siddharth Jha and Melissa Pan and Harshit Gupta and Parth Asawa and Carlos Guestrin and Matei Zaharia},
  title   = {Semantic Operators and Their Optimization: Enabling {LLM}-Based Data Processing with Accuracy Guarantees in {LOTUS}},
  journal = {Proc. VLDB Endow.},
  volume  = {18},
  number  = {11},
  pages   = {4171--4184},
  year    = {2025},
  doi     = {10.14778/3749646.3749685}
}

@inproceedings{palimpzest,
  author    = {Chunwei Liu and Matthew Russo and Michael Cafarella and Lei Cao and Peter Baile Chen and Zui Chen and Michael Franklin and Tim Kraska and Samuel Madden and Rana Shahout and Gerardo Vitagliano},
  title     = {{Palimpzest}: Optimizing {AI}-Powered Analytics with Declarative Query Processing},
  booktitle = {Proc. CIDR},
  year      = {2025},
  url       = {https://vldb.org/cidrdb/papers/2025/p12-liu.pdf}
}

@article{thalamusdb,
  author  = {Saehan Jo and Immanuel Trummer},
  title   = {{ThalamusDB}: Approximate Query Processing on Multi-Modal Data},
  journal = {Proc. ACM Manag. Data},
  volume  = {2},
  number  = {3},
  pages   = {1--26},
  year    = {2024},
  doi     = {10.1145/3654989}
}

@inproceedings{caesura,
  author    = {Matthias Urban and Carsten Binnig},
  title     = {{CAESURA}: Language Models as Multi-Modal Query Planners},
  booktitle = {Proc. CIDR},
  year      = {2024},
  url       = {https://www.vldb.org/cidrdb/papers/2024/p14-urban.pdf}
}

@inproceedings{aop,
  author    = {Jiayi Wang and Guoliang Li},
  title     = {{AOP}: Automated and Interactive {LLM} Pipeline Orchestration for Answering Complex Queries},
  booktitle = {Proc. CIDR},
  year      = {2025},
  url       = {https://www.vldb.org/cidrdb/papers/2025/p32-wang.pdf}
}

@inproceedings{unify,
  author    = {Jiayi Wang and Jianhua Feng},
  title     = {{Unify}: An Unstructured Data Analytics System},
  booktitle = {2025 IEEE 41st International Conference on Data Engineering (ICDE)},
  pages     = {4662--4674},
  year      = {2025},
  doi       = {10.1109/ICDE65448.2025.00374}
}

@article{nirvana,
  author  = {Junhao Zhu and Lu Chen and Xiangyu Ke and Ziquan Fang and Tianyi Li and Yunjun Gao and Christian S. Jensen},
  title   = {Beyond Relational: Semantic-Aware Multi-Modal Analytics with {LLM}-Native Query Optimization},
  journal = {Proc. ACM Manag. Data},
  volume  = {4},
  number  = {1},
  pages   = {14:1--14:25},
  year    = {2026},
  doi     = {10.1145/3786628}
}

@article{sembench,
  author  = {Jiale Lao and Andreas Zimmerer and Olga Ovcharenko and Tianji Cong and Matthew Russo and Gerardo Vitagliano and Michael Cochez and Fatma {\"O}zcan and Gautam Gupta and Thibaud Hottelier and H. V. Jagadish and Kris Kissel and Sebastian Schelter and Andreas Kipf and Immanuel Trummer},
  title   = {{SemBench}: A Benchmark for Semantic Query Processing Engines},
  journal = {Proc. VLDB Endow.},
  volume  = {19},
  number  = {8},
  pages   = {1754--1767},
  year    = {2026},
  doi     = {10.14778/3811243.3811249}
}

@article{tiinsight,
  author  = {Jun-Peng Zhu and Boyan Niu and Peng Cai and Zheming Ni and Jianwei Wan and Kai Xu and Jiajun Huang and Shengbo Ma and Bing Wang and Xuan Zhou and Guanglei Bao and Donghui Zhang and Liu Tang and Qi Liu},
  title   = {Towards Automated Cross-Domain Exploratory Data Analysis through Large Language Models},
  journal = {Proc. VLDB Endow.},
  volume  = {18},
  number  = {12},
  pages   = {5086--5099},
  year    = {2025},
  doi     = {10.14778/3750601.3750629}
}

@inproceedings{chat2query,
  author    = {Jun-Peng Zhu and Peng Cai and Boyan Niu and Zheming Ni and Kai Xu and Jiajun Huang and Jianwei Wan and Shengbo Ma and Bing Wang and Donghui Zhang and Liu Tang and Qi Liu},
  title     = {{Chat2Query}: A Zero-Shot Automatic Exploratory Data Analysis System with Large Language Models},
  booktitle = {2024 IEEE 40th International Conference on Data Engineering (ICDE)},
  pages     = {5429--5432},
  year      = {2024},
  doi       = {10.1109/ICDE60146.2024.00420}
}

\end{document}